\documentclass[epj]{svjour}
\input epsf
\begin{document}

\newcommand{\bc}{\begin{center}}
\newcommand{\ec}{\end{center}}
\newcommand{\be}{\begin{equation}}
\newcommand{\ee}{\end{equation}}
\newcommand{\beqn}{\begin{eqnarray}}
\newcommand{\eeqn}{\end{eqnarray}}

\newcommand{\DMRG}{{\sc dmrg}\ }    
\newcommand{\ARPACK}{{\sc arpack}\ }
\newcommand{\BST}{{\sc bst}\ }      
\newcommand{\FSM}{{\sc fsm}\ }      
\newcommand{\ISM}{{\sc ism}\ }      

\newcommand{\BEQ}{\begin{equation}}    
\newcommand{\BEA}{\begin{eqnarray}}
\newcommand{\EEQ}{\end{equation}} 
\newcommand{\EEA}{\end{eqnarray}}
\newcommand{\tr}{\widehat{\mbox{\rm tr\,}}}      
\newcommand{\eps}{\epsilon}       
\newcommand{\vph}{\varphi}        
\newcommand{\rar}{\rightarrow}    
\newcommand{\rcd}[1]{\bar{\cal D}_{#1}}          
\newcommand{\zeile}[1]{\vskip #1 \baselineskip}  
\newcommand{\vekz}[2]
     {\mbox{${\begin{array}{c} #1  \\ #2 \end{array}}$}}
\newcommand{\matz}[4]
     {\mbox{${\begin{array}{cc} #1 & #2  \\ #3 & #4 \end{array}}$}}

\title{Density Matrix Renormalization Group and Reaction-Diffusion
Processes}
\author{Enrico Carlon\inst{1}, Malte Henkel\inst{1} and
Ulrich Schollw\"{o}ck\inst{2}}
\institute{Laboratoire de Physique des Mat\'eriaux,\thanks{Unit\'e
Mixte de Recherche CNRS No~7556} Universit\'e Henri Poincar\'e Nancy I, 
B.P.~239, \\ F - 54506 Vand{\oe}uvre l\`es Nancy Cedex, France
\and
Sektion Physik, Ludwig-Maximilians-Universit\"at M\"unchen,
Theresienstr. 37/III, D - 80333 M\"unchen, Germany}
\date{29$^{\rm th}$ of January, 1999}
\abstract{
The density matrix renormalization group ({\sc dmrg}) is applied to
some one-dimensional reaction-diffusion models in the vicinity
of and at their critical point. The stochastic time evolution for these
models is given in terms of a non-symmetric ``quantum Hamiltonian'', which is
diagonalized using the \DMRG method for open chains of moderate lengths (up to
about $60$ sites). The numerical diagonalization methods for non-symmetric
matrices are reviewed. Different choices for an appropriate density matrix in
the non-symmetric \DMRG are discussed. Accurate estimates of the steady-state
critical points and exponents can then be found from finite-size scaling
through standard finite-lattice extrapolation methods. This is exemplified
by studying the leading relaxation time and the density profiles of
diffusion-annihilation and of a branching-fusing model in the directed
percolation universality class.
} 

\authorrunning{E. Carlon, M. Henkel and U. Schollw\"ock}
\titlerunning{Density Matrix Renormalization Group and 
Reaction-Diffusion Processes}
\maketitle
\section{Introduction}
\label{intro}

The density matrix renormalisation group ({\sc dmrg}) technique was
invented by White \cite{Whit92} in 1992 as a new tool for the
diagonalization of quantum chain spin Hamiltonians. The \DMRG allows 
to study much larger systems than is possible with standard exact
diagonalization methods and provides data with remarkable 
accuracy. The method has since been applied to a large variety of 
systems in quantum \cite{quantum} and classical \cite{classical} physics.
These applications usually considered hermitian quantum Hamiltonians 
for spin chains or symmetric transfer matrices for two-dimensional
classical systems. In both cases the \DMRG is known to work very well.
For a collection of reviews see \cite{DMRGBook}. 

Motivated by the success obtained for these problems, some efforts have 
been recently devoted to problems involving diagonalization of non-symmetric 
matrices which are often encountered in various fields of physics; as 
example we mention the cases of low temperature thermodynamics of spin 
chains \cite{bursill} or out-of-equilibrium classical systems 
\cite{hieda,kaulke}. 

The models investigated in this paper belong to the latter class of
problems. We consider one-dimensional reaction-diffusion systems 
where the stochastic time evolution of the system is given 
by a non-hermitean operator $H$ to which we apply the \DMRG algorithm.
One of the models considered displays a non-equilibrium phase transition
which is expected to be in the universality class of 
directed percolation, a typical class for many out-of-equilibrium
systems and for which the critical exponents are known numerically
to a high degree of accuracy.

Our main motivation for this study is as follows. At equilibrium, studies of 
finite systems through 
diagonalization of a transfer matrix and/or a quantum Hamiltonian and analysis
of the results through finite-size scaling, eventually combined with
precise extrapolation algorithms, has been one of the standard methods of
studying phase transitions in systems with many strongly interacting degrees
of freedom \cite{Priv90,Card88,Henk99}. 
Compared to more standard diagonalization methods, the \DMRG does
allow to study much larger systems, even if it works best in situations
when the ground state of $H$ is 
separated by a finite gap from the first excited 
state, that is, at a finite distance away from the critical point. That 
advantage is at least partially offset by the fact that for reasons of
numerical accuracy, it is preferable to apply the \DMRG to 
models with {\em open}
boundary conditions, which lead to estimates of the critical parameters which
converge usually slower than those found for periodic boundary conditions. 
Here we study to what extent a finite-size diagonalization study of
{\em non-equilibrium} systems is feasible and in particular, to what degree
of precision critical points and critical exponents can be estimated. For that
purpose, we shall consider some critical non-equilibrium models with known
properties to compare with our \DMRG data. 

Diagonalizing non-symmetric matrices is numerically much more 
demanding than for symmetric matrices. Potentially, numerical instabilities
might arise at several stages of the calculation. We shall discuss
in some detail various diagonalization methods for non-symmetric matrices and 
have tested our results throughout by working with two different 
diagonalization algorithms.

We shall not only consider the calculation of eigenvalues, which determine
the relaxation times, but also of eigenvectors, which are needed in the 
calculation of matrix elements, as they arise for example in the
calculation of density profiles.
In general, we find that for non-symmetric matrices the
numerical accuracy is not as good as for symmetric ones. However, we shall
show that the \DMRG is well capable to accurately determine 
the values of the critical parameters. This makes this technique a 
useful general-purpose method for the study of non-equilibrium 
critical phenomena.

This paper is organized as follows: in section \ref{sec:rdproc}, we 
introduce reaction-diffusion systems and their description in terms 
of non-symmetric evolution matrices, in section \ref{sec:dmrg} we 
review the basics of the \DMRG algorithm, and present in section 
\ref{sec:diag} a brief review of diagonalization methods for 
non-symmetric matrices.
More specific details about choosing a convenient density matrix for 
non-equilibrium systems are described in section \ref{sec:choice}. 
In sections \ref{sec:fss} and \ref{sec:densityprof} we present our 
numerical results concerning the critical properties of the model 
and then conclude. 

\section{Reaction-diffusion processes}
\label{sec:rdproc}

We consider a chain of length $L$ in which each site is either occupied 
by a particle ($A$) or empty ($\emptyset$). The time evolution of the 
system is given in terms of microscopic rules involving only a pair of 
particles on neighbouring sites. We shall consider the following system,
using the notation of \cite{Henk97}:
\begin{eqnarray}
A\emptyset \leftrightarrow \emptyset A \,\,\,\,
({\rm with}\,\,{\rm rate}\,\,D)
\label{diff}\\
AA \rightarrow  \emptyset \emptyset  \,\,\,\,\,
({\rm with}\,\,{\rm rate}\,\,2\alpha) 
\label{pairannih}\\
AA \rightarrow  \emptyset A, A\emptyset  \,\,\,\,\,
({\rm with}\,\,{\rm rate}\,\,\gamma) 
\label{coagul}\\
A\emptyset , \emptyset A \rightarrow  \emptyset \emptyset  
\,\,\,\,\,({\rm with}\,\,{\rm rate}\,\,\delta)
\label{death}\\
A\emptyset , \emptyset A \rightarrow  AA \,\,\,\,\,
({\rm with}\,\,{\rm rate}\,\,\beta)
\label{decoag}
\end{eqnarray}
Apart from the diffusion process (\ref{diff}), all the other processes 
involve either a decrease (\ref{pairannih},\ref{coagul},\ref{death}) 
or an increase (\ref{decoag}) in the number of particles (see \cite{Henk97}
for a list of common alternative notations).  

Once the reaction rates are given, the stochastic evolution follows from
the master equation, which can be written as 
\begin{equation}
\frac{d | P(t) \rangle }{d t} = - H | P(t) \rangle
\label{maitresse}
\end{equation}
where $| P(t) \rangle$ is the state vector. The elements of the 
quantum ``Hamiltonian'' $H$ are given by
\BEA
\langle \sigma | H | \tau \rangle &=& - w(\tau \rightarrow \sigma) \;\; ; \;\;
\sigma\neq\tau \nonumber \\
\langle \sigma | H | \sigma \rangle &=& \sum_{\tau \neq \sigma}
w( \sigma \rightarrow \tau)
\EEA
where $| \sigma \rangle$, $| \tau \rangle$ are the state vectors of the
particle configurations $\sigma,\tau$ and
$w(\tau \rightarrow \sigma)$ denotes the transition probability 
between the two states and is easily constructed from the rates 
(\ref{diff}-\ref{decoag}) of the elementary processes. Since $H$ is a
stochastic matrix, the left ground state $\langle s|$ is
\begin{equation}
\langle s| = \sum_{\sigma} \langle \sigma |
\end{equation}
with ground state energy $E_0=0$, since $\langle s|H =0$. All other eigenvalues 
$E_i$ of $H$ have a non-negative real part $\Re E_i \geq 0$ \cite{Kamp81}. 
Since the formal solution of eq.~(\ref{maitresse}) is
\BEQ
|P(t)\rangle = e^{-Ht} |P(0)\rangle
\EEQ
the system evolves towards its steady state $|P(\infty)\rangle$. Let
$\Gamma := \inf_{i} \Re E_i$ for $i\neq 0$. 
Often, one simply has 
\BEQ
\Gamma=E_1-E_0 = E_1
\EEQ
since $E_0=0$. If
$\Gamma>0$, the approach towards the steady state is characterized by a 
{\em finite}\ relaxation time $\tau=1/\Gamma$, but if $\Gamma=0$, that
approach is algebraic. This situation is quite analogous to non-critical
phases ($\tau\neq 0$) and critical points ($\tau=0$), 
respectively, which may arise in equilibrium quantum Hamiltonians. 

It is clear that $H$ is non-symmetric if $w( \sigma \rightarrow 
\tau) \neq w(\tau \rightarrow \sigma)$. However, if the detailed-balance
condition
\BEQ
w(\sigma\rar\tau) P_{s}(\{\tau\}) = w(\tau\rar\sigma) P_{s}(\{\sigma\})
\EEQ
is satisfied, where $P_{s}(\{\sigma\})$ is defined by \linebreak
$|P(\infty)\rangle = \sum_{\sigma} P_{s}(\{\sigma\}) |\sigma\rangle$, $H$ is
similar to a non-stochastic, but {\em symmetric} matrix $K$, without affecting
the locality of the interactions, e.g. \cite{Kamp81,Ciep98}. Detailed balance
always holds when besides diffusion only a {\em single} reversible reaction
is present. For several reversible reactions, the cases when detailed balance
holds are given in \cite{Alca94}, eq.~(4.48). Only if detailed balance holds,
the right ground state $|s\rangle=|P(\infty)\rangle$ is related to the known 
left ground state $\langle s|$ in a simple way \cite{fuss1}.  

In this paper, we shall study systems {\em without}\ detailed balance, since 
it is our aim to explore the \DMRG in a setting as different from equilibrium
physics as possible. Specifically, we shall consider two models:

(1) {\em Diffusion-annihilation.} The rates are
\BEQ
D = 2 \alpha = p \;\; , \;\; \beta=\gamma=\delta=0
\label{da}
\EEQ
This model is characterized by an algebraic approach towards the steady 
state with a gap vanishing in the thermodynamic limit as $\Gamma \sim 
1/L^2$. It offers to us the additional 
advantage that $H$ can be diagonalised through free-fermion methods (e.g. 
\cite{Alca94}) and thus analytical results are available for comparison with 
the numerical data. 

(2) {\em Branching-fusing.} Here we take
\BEQ
D=2\alpha=\gamma=\delta=1-p \;\; , \;\; \beta=p
\label{bf}
\EEQ
This model presents a non-equilibrium phase transition. For $p$ small, the
annihilating processes dominate and the steady state is simply the empty 
lattice. However, if $p$ becomes sufficiently large, the steady state contains
particles at some finite density. The transition between these two phases
is expected to fall into the directed percolation universality class
(see also appendix~\ref{sec:appendixa}). 
Although even in $1D$, there is no analytical information available, series
expansion studies and Monte Carlo simulation have over the years yielded
extremely precise estimates of critical exponents, as reviewed in \cite{Essa96}
and \cite{Muno98,Laur97}. We collect some of their 
results in table~\ref{MCexp}, but
skip over repeating the precise exponent definitions here.
We use the notation $\theta=\nu_{\|}/\nu_{\perp}$ for the dynamical exponent
in order to avoid confusion with the exponent $z=2/\theta$ which is also often 
used. If $D=0$, the model would reduce to standard site-bond directed 
percolation, but the presence of $D$ should not change the universality class. 
The last column of table \ref{MCexp} shows the values of the exponents
obtained from finite-size scaling extrapolation of our numerical data. 

\begin{table}
\caption{Some critical exponents for $1D$ directed percolation, as obtained
from series data {\protect \cite{Essa96}} and Monte Carlo simulations 
{\protect \cite{Muno98,Laur97}}. In the last column, the 
values obtained by density matrix renormalization in the present work are 
listed. The numbers in brackets give the estimated uncertainty.} 
\begin{tabular}{|c|c|c|c|} 
\hline
exponent                      & series      & simulation & this work\\ \hline
$\beta$                       & 0.27647(10) & 0.27649(4) &           \\
$\beta/\nu_\perp$             & 0.2520(1)   & 0.25208(4) & 0.249(3) \\
$\beta_1$                     & 0.7338(1)   & 0.728(1)   &           \\
$\beta_1/\nu_\perp$           & 0.6690(1)   & 0.664(7)   & 0.667(2) \\
$\nu_{\|}$                    & 1.7338(1)   & 1.73383(3) &         \\
$\nu_{\perp}$                 & 1.0969(1)   & 1.09684(1) & 1.08(2)  \\
$\theta=\nu_{\|}/\nu_{\perp}$ & 1.5806(2)   & 1.58074(4) & 1.580(1) \\ 
\hline
\end{tabular} 
\label{MCexp}
\end{table}

In both models (\ref{da},\ref{bf}),
the empty lattice $|\emptyset\ldots\emptyset\rangle$ is
obviously an adsorbing state and consequently also a steady state.
Since $H$ is real, its eigenvalues are either real or occur in complex
conjugate pairs. The first excited eigenvalue $E_1$ is always real.

For the branching-fusing model, we also considered the case of particle
injection at the boundaries of the system, i.e. we added the reaction
\beqn
\emptyset  \rightarrow  A \,\,\,\,\,({\rm with}\,\,{\rm rate}\,\,p')
\label{inject}
\eeqn
at the two sites at the edges.
In this situation, $|\emptyset\ldots\emptyset\rangle$ is no longer
a stationary state of the system since the lattice is occupied by
a non-vanishing density of particles for all values of $p$. We analyze
the shape of the density profiles as function of the parameter $p$
and of the injection rate $p'$.

\section{Density matrix renormalization group algorithm}
\label{sec:dmrg}

We recall briefly in this section the basic structure of the \DMRG
algorithm \cite{Whit92}. The task is to
find selected approximate eigenvalues and eigenvectors of
a given Hamiltonian $H$. That desired eigenvector $|\psi\rangle$
is called a {\em target state} and the process of selecting $|\psi\rangle$ is
referred to as {\em targeting} (it is possible to target several states, see
below). One assumes that $H$ as defined on a
open chain with $L$ sites and has the local structure
\BEQ
H = \sum_{i=1}^{L-1} h_{i,i+1}
\label{ham}
\EEQ
where $h_{i,i+1}$ is a local Hamiltonian acting on a pair of nearest-neighbour
sites (this condition of locality can be somewhat relaxed but we restrict here
to the simplest case). In this paper, we always consider free (open) boundary
conditions, as typically done in \DMRG studies.

The \DMRG is an iterative method: it produces a chosen eigenvector (usually,
one targets the ground state or the first excited state) and its eigenvalue,
starting from the target state of a small chain which is known
from exact diagonalization of the
Hamiltonian (\ref{ham}), and then using it to find $|\psi\rangle$ on chains
with an increasing number of sites.
This is made possible by projecting at each iteration step
the full vector space
to a smaller space where only a selected number $m$ of states is kept.
This projection is carried out via the density matrix
as described below.

A \DMRG calculation, at least in the context of applications to
critical phenomena, proceeds in two steps.
The first one is the infinite system method ({\sc ism}) which
we now describe. Suppose we are interested in the ground state of $H$.
As the starting point, consider a chain of four lattice sites
which can be represented as $B_l^{(1)} \bullet \bullet B_r^{(1)}$, where
$\bullet$ denotes a single site and $B_{r,l}^{(1)}$ are blocks at the left and
right side of the chain. Initially, they contain only one spin, that is
$B_{r,l}^{(1)} = \bullet$ (of course, the calculation may be started at larger
lattices)

At this point, the main loop begins.
The Hamiltonian $H$ is
easily written down and its ground state wave function
$\psi_0 (\alpha_l , i_l , j_r , \beta_r )$ can be found via standard
diagonalization routines, where
$\alpha_l$ and $\beta_r$ denote degrees of freedom of the blocks $B_{r}^{(1)}$
and $B_{l}^{(1)}$ and the indices
$i_l , j_r $ refer to the spin degrees of freedom of the single lattice
points in the middle of the chain.
The density matrix for the left part of the system is defined as
\be
\rho^{(l)} (\alpha_l , i_l ; \gamma_l , k_l ) =
\sum_{j_r , \beta_r }
\psi_0 (\alpha_l , i_l , j_r , \beta_r )
\psi_0 (\gamma_l, k_l , j_r , \beta_r )
\label{definedm}
\ee
which we shall write in a short-hand notation as
$\rho = \tr ( | \psi_0 \rangle \langle \psi_0 | )$, where $\tr$ denotes a
partial trace either in the left or right part of the system (we shall
reconsider the precise choice of $\rho$ in section \ref{sec:choice}).
Next, one solves the eigenvalue problem $\rho|\Omega_i\rangle =\omega_i|
\Omega_i\rangle$.
The eigenvalues of the density matrix are non-negative
and can be ordered according to $\omega_1 \geq \omega_2 \geq \omega_3
\geq \ldots \geq 0$. Furthermore,
if the ground state vector of $H$ is normalized
according to $\langle \psi_0 | \psi_0 \rangle =1$, one has
$\sum_i \omega_i = 1$. Each eigenvalue $\omega_i$ is equal to the probability
of finding the left part of the chain in the corresponding density matrix
eigenvector $| \Omega_i \rangle$ when the whole system is in the ground
state $| \psi_0 \rangle$.
The configurational space reduction is obtained by keeping only the first
$m$ dominant density matrix eigenvectors $| \Omega_i \rangle $ with $i =
1,2, \ldots m$, corresponding to the $m$ largest $\omega_i$.
Formally, the truncation can be represented by
\be
O_m^T \left( B_l^{(1)} \bullet \right) O_m = B_l^{(2)}
\label{projection}
\ee
where $O_m =[|\Omega_1\rangle, \ldots ,|\Omega_m\rangle]$. The
accuracy of the projection operation can be described by the truncation error:
\be
\epsilon = 1 - \sum_{i=1}^{m} \omega_i
\ee
The projection operation is repeated for the right part as well to
obtain $B_r^{(2)}$ (if there is a left-right symmetry, as in the models
studied here, $B_r^{(2)}$ is obtained simply from reflection of $B_l^{(2)}$).
Performing these calculations,
sparse-matrix diagonalization techniques
will be needed to obtain the ground eigenstate $|\psi_0\rangle$ of $H$.
On the other hand, since the whole spectrum of the non-sparse matrix
$\rho$ is required, this is best obtained via some standard routines.

Combining the two blocks with new sites one gets $B_l^{(2)} \bullet
\bullet B_r^{(2)}$, e.g. a chain of $L=6$ sites after the first pass through
the main loop. The next pass through the main loop begins by writing down
$H$ for this longer chain.

Applying this procedure repeatedly at the left and right part of the
system, one generates larger systems. At each iteration step, two new
sites are added in the middle of the chain and the boundaries are pushed
further away from each other. Schematically, this may be illustrated as
\begin{displaymath}
B_l^{(1)} \bullet \bullet B_r^{(1)}  \rightarrow
B_l^{(2)} \bullet \bullet B_r^{(2)} \rightarrow  \ldots
\rightarrow B_l^{(L/2 - 1)} \bullet \bullet B_r^{(L/2 - 1)}
\end{displaymath}
The {\sc ism} procedure is repeated,
typically until $L \approx 1000$, and if there is a finite gap in the
low-lying spectrum of $H$, finite-size effects can then be neglected.
However, it is well known that the {\sc ism} method alone is not
enough to yield precise results in the thermodynamic limit $L \to\infty$
for systems close to criticality \cite{ostlund,DMRGBook}.

In such cases (e.g. \cite{ferenc}), a second step in the \DMRG calculation
is required. It is best to use the finite
system method ({\sc fsm}) designed by White \cite{Whit92} to accurately
determine properties of systems of finite length. The starting point of the
{\sc fsm} is the target vector $|\psi\rangle$ for a chain of given length
$L$, as generated by the {\sc ism} described above.
At this point further iterations are started. First, one calculates better
approximations for the blocks on the
left part representing more than $L/2-1$ sites, using as before
eq. (\ref{projection}), while for the blocks on the right part, one
uses blocks generated in previous iterations in order to keep the total
length of the system fixed at $L$. Schematically this looks as follows
\BEA
B_l^{(L/2 - 1)} \bullet \bullet B_r^{(L/2 - 1)} \rightarrow
B_l^{(L/2)} \bullet \bullet B_r^{(L/2 - 2)} \rightarrow
\nonumber \\
\ldots \rightarrow B_l^{(L - 3)} \bullet \bullet B_r^{(1)} \nonumber
\EEA
Second, this procedure is reversed and the larger blocks on right part of the
system are refined. Schematically,
\BEA
B_l^{(L - 3)} \bullet \bullet B_r^{(1)} \rightarrow
B_l^{(L - 2)} \bullet  \bullet B_r^{(2)} \rightarrow 
\ldots \rightarrow B_l^{(1)} \bullet \bullet B_r^{(L - 3)}
\nonumber
\EEA
In these steps, on the $B_{r}$ are updated according to eq.~(\ref{projection}),
while the $B_{l}$ are taken from the blocks calculated previously. Finally,
the $B_l$ are updated again through the sequence
\BEA
B_l^{(1)} \bullet \bullet B_r^{(L - 3)} \rightarrow
B_l^{(2)} \bullet \bullet B_r^{(L - 4)} \rightarrow
\nonumber \\
\ldots \rightarrow B_l^{(L/2 - 1)} \bullet \bullet B_r^{(L/2 - 1)} \nonumber
\EEA
until one is back at the left-right symmetric partition. The target vector
extracted at this stage can be used as starting point for the next {\sc fsm}
iteration.

These ``sweeps'' improve the results
both for eigenvalues and eigenvectors \cite{Whit92}. For a given lattice
size, two or three sweeps are usually enough to achieve convergence.

In practice, to calculate critical exponents, one needs data
for chains of various lengths in order to perform a finite-size scaling
analysis. For better efficiency, we used repeatedly the {\sc fsm}
to calculate from the same
run quantities for chains of different lengths as follows. The blocks
generated at the end of the {\sc fsm} for a chain of length $L_0$ are
used as starting point for further \DMRG calculation: first, we use the
{\sc ism} to enlarge the system is
symmetrically until a length $L_1> L_0$ is reached. Second, the
{\sc fsm} sweeps are started again. In this way we produced accurate results
for systems of various lengths $L_0 \leq L_1 \leq \ldots$, typically equally
spaced, in the same \DMRG run.

\section{Diagonalization methods for non-symmetric matrices}
\label{sec:diag}

In ``conventional''
\DMRG calculations the computationally dominant part is the
determination of (ground state) wave functions
given by the eigenvectors of a large sparse symmetric
$N \times N$ matrix $A$ using a diagonalization algorithm for
large sparse symmetric matrices such as the Lanczos algorithm.
Using the fundamental operation
\begin{equation}
|w\rangle = A|v\rangle ,
\end{equation}
one builds iteratively small tridiagonal
$n\times n$ matrices $T_{n} = Q^{T}_{n}AQ_{n}$ with
$n \ll N$, where $Q_{n}$ is a $N \times n$ matrix with orthonormal columns.
The matrices $T_{n}$ are diagonalisable by standard techniques; it can
be shown that their extreme eigenvalues converge for rather small $n$ to
the extreme eigenvalues of $A$. Moreover,
their eigenvalues form Sturm chains: if $\lambda_{i}$ are the eigenvalues
of $T_{n}$ in ascending order and $\mu_{i}$ those of $T_{n+1}$, one has
$\mu_i < \lambda_{i} < \mu_{i+1}$. This ensures monotonous, easily
controlled convergence of the extreme eigenvalues. $Q_{n}$ relates the
eigenvectors of $T_{n}$ to those of $A$: $|\lambda_{0}\rangle_{A} =
Q_{n}|\lambda_{0}\rangle_{T_{n}}$.
In the case of large sparse non-symmetric matrices, as they arise in
transfer matrix \DMRG and non-hermitian {\sc dmrg}, the situation is much
less satisfying, as the lack of symmetry leads to intrinsic problems
of numerical stability.

Essentially, there are two algorithms available, the Ar\-nol\-di algorithm
and the non-symmetric Lanczos algorithm \cite{Golub}. In our calculations,
we used both methods, to check them against each other to guard against
numerical fallacies. Usually, they were in excellent agreement, though the
professional package available for the Arnoldi algorithm \cite{ARPACK}
usually seemed to be somewhat more accurate than our self-made
implementation of the non-symmetric Lanczos algorithm.

Both also build on the fundamental
operation
\begin{equation}
|w\rangle = A|v\rangle
\end{equation}
and its transpose
\begin{equation}
\langle w| = \langle v| A
\end{equation}
in the case of the non-symmetric Lanczos algorithm.
Again, one builds iteratively small $n\times n$ matrices with
$n \ll N$ diagonisable by standard techniques
for small non-symmetric matrices. In both cases, one finds that the extreme
eigenvalues of the small matrices do not converge
to those of the big matrices in the very systematic
manner of the symmetric case, i.e.\ they do not form Sturm chains.

The Arnoldi method iteratively generates one sequence of
orthonormal vectors $|q_{i}\rangle$
forming the columns of a matrix
$Q_{n}=[|q_{1}\rangle,\ldots,|q_{n}\rangle]$
and a Hessenberg matrix $H_{n} = Q^{T}_{n}AQ_{n}$ with elements
$h_{ij}$. One starts from a
random vector $|q_{0}\rangle$ of unit length and $h_{10}=1$ and iterates
using
\begin{equation}
h_{k+1,k}|q_{k+1}\rangle = A|q_{k}\rangle - \sum_{i=1}^{k} h_{ik}|q_{i}\rangle
\end{equation}
with $h_{ij}=\langle q_{i}| A | q_{j}\rangle$. $h_{k+1,k}$ is determined
by enforcing unit length for $|q_{k+1}\rangle$. Here,
$\langle q_i| = |q_i\rangle^{T}$.

The non-symmetric Hessenberg
matrix is then diagonalized using the standard QR algorithm. Some of the
eigenvalues of $H_{n}$ will converge against some of those of $A$,
and associated eigenvectors of $H_{n}$ can be transformed into those of $A$
using $Q_{n}$. However, it is quite intricate to assure that eigenvalue
convergence actually happens numerically.
We will not discuss this issue further,
as there is a freely available highly sophisticated
Arnoldi package ({\sc arpack})\cite{ARPACK}.

Let us now discuss the non-symmetric Lanczos
algorithm, which is quite easily implemented, but has to be carefully
refined for numerical stability. Then it provides a highly satisfying
alternative approach portable to machines where \ARPACK might not be available.

Let us first
summarize the algorithm as presented in Ref. \cite{Golub}.
The non-symmetric Lanczos algorithm generates iteratively {\em two} sequences
of vectors from whom eventually left and right eigenvectors are built.
In an adaptation of the symmetric case, one forms matrices from these vectors,
$Q_{n} = [|q_{1}\rangle,\ldots,|q_{n}\rangle]$ and
$P_{n} = [\langle p_{1}|,\ldots,\langle p_{n}|]$, to generate an (incomplete)
basis transformation from the $N\times N$ matrix $A$ to a tridiagonal
$n\times n$ matrix $T_{n}$,
\begin{equation}
T_{n} = P_{n} A Q_{n} = \left[ \begin{array}{ccccc}
\alpha_{1} & \gamma_{1} & . & . & 0 \\
\beta_{1}  & \alpha_{2} & \gamma_{2} & . & . \\
. & . & . & . & . \\
. & . & . & . & \gamma_{n-1} \\
0 & . & . & \beta_{n-1} & \alpha_{n}
\end{array} \right]
\end{equation} 
and
demands {\em biorthogonality}, $P_{n}Q_{n}=I_{n}$. One then obtains
the following equations determining the
elements of the tridiagonal matrix:
\begin{eqnarray}
\beta_{k} |q_{k+1}\rangle &:=& |t_{k}\rangle = (A - \alpha_{k}I)
| q_{k} \rangle - \gamma_{k-1} | q_{k-1} \rangle \\
\gamma_{k} \langle p_{k+1}| &:=& \langle s_{k}| =
\langle p_k | (A - \alpha_{k}I) - \beta_{k-1} \langle p_{k-1} |
\end{eqnarray}
with
\begin{eqnarray}
\alpha_{k} &=& \langle p_{k} | A | q_{k} \rangle \\
\beta_{k} &=& ||\, | t_{k} \rangle || \\
\gamma_{k} &=& \langle s_{k} | t_{k} \rangle /\beta_{k} .
\end{eqnarray}
Note that the choice of the off-diagonal coefficients is not unique.
One starts with $|q_{0}\rangle = 0$ and
$\langle p_{0}| = 0$ and two non-orthogonal random vectors $|q_{1}\rangle$
and $\langle p_{1}|$, normed such that $|q_{1}\rangle$ has unit length and
$\langle p_{1} | q_{1} \rangle=1$ as input for the first iteration.
This leads to an iteratively growing $T_{n}$. The
iterations are continued, until the lowest (two) eigenvalues of $T_{n}$
are sufficiently converged. From the left and right
eigenvectors of $T_{n}$ one forms
the left and right eigenvectors of $A$: $|\lambda\rangle_{A} =
Q_{n}|\lambda\rangle_{T_{n}}$ and
$\langle\lambda|_{A} = \langle\lambda|_{T_{n}}P_{n}$.

Note that the non-symmetric Lanczos algorithm yields left and right
eigenvectors on an equal footing, while the Arnoldi algorithm has to be run
twice for $A$ and $A^{T}$ (however, with less time-consuming matrix-vector
multiplications).

In the present form, the non-symmetric Lanczos
algorithm is capable of yielding rather good
eigenvalues. However, eigenvectors are determined with rather poor accuracy
only. This leads to a poor choice for the density matrix,
in turn to a non-optimized decimation and then, after some \DMRG steps, to
a noticeable degradation of the eigenvalues too. There are two reasons
for this.

(1) The construction of the two Lanczos vector sequences guarantees them
to be biorthogonal, or $\langle p_{i} | q_{j} \rangle = \delta_{ij}$.
While true mathematically, this biorthogonality is numerically only true
locally, i.e.\ if $i$ and $j$ are close. Globally, small overlaps
develop, in particular when the extreme eigenvalues of $T_{n}$
start to converge.
This loss of orthogonality introduces numerical errors in the mapping
from the eigenvectors of $T_{n}$ to those of $A$. An analogous loss of
orthogonality occurs in the Arnoldi algorithm.

(2) Even the symmetric Lanczos algorithm shows the phenomenon that if one
wants a non-extreme, say the second, eigenvalue to converge, it may
occur that the eigenvalue converges to its true value, but if one
enforces too strict convergence criteria, ``jumps'' and will converge
to the lowest eigenvalue, too, see \cite{Henk99} for an example. 
This phenomenon is much more frequent
in the non-symmetric Lanczos algorithm
and because of the non-monotonic convergence behaviour
much harder to detect and handle numerically. One therefore has to relax
convergence criteria, which lead to a not so good eigenvalue, but in
particular a rather bad eigenvector.

Both issues can be addressed in a satisfactory way.
Basically, all Lanczos methods are clever implementations of the
power method, which obtains the largest eigenvector of a matrix
by applying $A$ successively to a random start vector $|v\rangle$:
for $n \rightarrow \infty$, $|\lambda_{max}\rangle = A^{n}|v\rangle /
\| A^{n}|v\rangle \|$. The
disadvantage of the power
method is its very slow convergence; however, it
accumulates no numerical inaccuracies from previous iterations: it essentially
restarts at each iteration with a better guess for the eigenvector.
A power iteration is thus a suitable eigenvector ``beautifier'': once
Lanczos has effectively converged, it can be used to improve the
generated eigenvector (and also somewhat the eigenvalue) by eliminating
accumulated numerical inaccuracies. As we try to improve the eigenvectors
of the smallest eigenvalues, we apply the power method to $A \rightarrow -A
+ (\mbox{tr\ }A)I$, turning the smallest eigenvalues into the largest by
absolute value, as the real parts of all eigenvalues are non-negative, while
conserving the eigenvectors.
As we have a left and right eigenvector on equal footing, one has to
slightly modify the power method as follows: Let $|r_{0}\rangle$ and
$\langle l_{0}|$ be the right and left eigenvectors obtained by Lanczos
for the smallest (now largest) eigenvalue.
One now forms successively $|r\rangle = A|r_{0}\rangle$,
$\lambda = \langle l_{0} | r \rangle / \langle l_{0}|r_{0}\rangle$
as improved eigenvalue, and
sets $|r_{1}\rangle = |r\rangle / \lambda$ as improved right eigenvector.
Now one forms $\langle l| = \langle l_{0}| A$,
$\lambda = \langle l | r_{1} \rangle / \langle l_{0}|r_{1}\rangle$
as improved eigenvalue,
and sets $\langle l_{1}| = \langle l| / \lambda$ as improved left
eigenvector,
and restarts with the new eigenvector pair, until the
eigenvalue is satisfactorily converged. We find empirically that if we apply
this procedure to the second eigenvalue in the spectrum there
is no problem that it will start to
converge to
the first one for a small number of iterations. It is therefore suitable
to improve both the lowest (largest)
and the second eigenvalue and the associated
eigenvectors.

\section{The choice of the density matrix}
\label{sec:choice}

An important question to address in the case of a \DMRG calculation for a
non-symmetric problem concerns the choice of density matrix. For reference,
we recall the situation found in studies of equilibrium
quantum spin chains at a non-zero temperature. There,
a non-symmetric transfer matrix
is generated by applying a Trotter decomposition along the
imaginary time direction. The density matrix typically used in these
systems is defined by \cite{bursill}
\be
\rho = \tr  \left\{ |\psi^{(l)}_0 \rangle \langle
\psi^{(r)}_0|\right\}
\label{dmqsc}
\ee
where $\tr$ denotes the partial trace on part of the system, the
superscripts $l$ and $r$ label the left and right eigenvectors of the
transfer
operator $\tilde{T}$ (throughout this section, the line vector
$\langle \psi^{(r)}_0|$ is the transpose of the column vector
$|\psi^{(r)}_0 \rangle$ and so on).
The choice (\ref{dmqsc}) is justified as the one
that maximizes the partition function which describes the thermodynamics
of the system.
Since $\rho$ is non-symmetric, it may have complex eigenvalues.
In practice, however, non-real eigenvalues
appear only after a certain number of
\DMRG steps. They are thought to come from numerical round-off errors.
Several approaches have been invented to avoid these problems
\cite{bursill}.

In the present study, we restrict ourselves to symmetric density
matrices. Usually, convergence was best for the type
\BEQ
\rho_i^{[1]} := \frac 1 2 \,\, \tr \left\{ |\psi^{(l)}_i \rangle
\langle \psi^{(l)}_i| + |\psi^{(r)}_i \rangle \langle
\psi^{(r)}_i| \right\}
\label{densitymat}
\EEQ
where $|\psi^{(l)}_i \rangle$ and $|\psi^{(r)}_i \rangle$
denote the (normalized) left and right eigenvectors corresponding
to the $i$-th eigenvalue of the non-symmetric Hamiltonian defined
in (\ref{maitresse}).
This choice is easy to implement numerically,
since one avoids all problems related to the possibility of
complex eigenvalues of a non-symmetric density matrix, but has
also a more profound justification. Recall
that in White's argument \cite{Whit92}, the choice of
density matrix does not rely on the Hamiltonian being symmetric.
For a symmetric Hamiltonian the density matrix
obtained from the combination $|\psi_0 \rangle \langle \psi_0|$
allows the construction of a trial ground state function $\tilde
{|\psi_0 \rangle }$ whose distance from the exact ground state
$|\psi_0 \rangle$ is minimal \cite{Whit92}.
In the non-symmetric case, the density matrix defined by
eq. (\ref{densitymat}) provides a basis set
which minimizes simultaneously the distance of the trial
vectors from the exact right and left eigenstates
$|\psi^{(l)}_i \rangle$ and $|\psi^{(r)}_i \rangle$,
see appendix~\ref{sec:appendixb}.

We also tried out some alternatives for a symmetric density matrix.
For example, one might consider $\rho$ being defined
by using the right eigenstate only
\be
\rho_i^{[2]} := \tr \left\{ |\psi^{(r)}_i \rangle
\langle \psi^{(r)}_i| \right\}
\label{densitymatright}
\ee
(this density matrix was used in a study of the $q$-symmetric Heisenberg
chain \cite{kaulke}).
Alternatively, one can use a density matrix with mixed terms
\be
\rho_i^{[3]} :=  \alpha_{i} \,\,\, \tr \left[ \left(
| \psi_i^{(l)} \rangle + | \psi_i^{(r)} \rangle
\right) \left( \langle \psi_i^{(l)} | + \langle
\psi_i^{(r)} | \right) \right],
\label{densitymatmix}
\ee
with $\alpha_{i}$ a normalization constant. The merit of this density
matrix would be that, while keeping the advantages of being symmetric,
it contains terms $| \psi_i^{(l)} \rangle \langle \psi_i^{(r)} |$ which
holds the relevant information in the case of \DMRG quantum thermodynamics.

Which of the density matrices $\rho^{[1]}$ (\ref{densitymat}),
$\rho^{[2]}$ (\ref{densitymatright}) and $\rho^{[3]}$ (\ref{densitymatmix})
is the `best' one, depends on the Hamiltonian under study and also of
the properties of the eigenstates one wants to calculate.
This choice is {\it a priori} a rather difficult one.
Our aim is to select the density matrix that produces the most stable
numerical results. How to do this is best illustrated by some examples.

Table \ref{densitymatrices}
shows the finite system method ({\sc fsm}) iterations for the
branching-fusing model defined in eq. (\ref{bf}) for a chain of length
$L=20$, with $m=32$ states kept and with $p=0.84$, thus in the
vicinity of the critical point (we estimate $p_c = 0.84036(1)$ in the next
section). The data were obtained by using only the first excited state as a
target state. The $L=20$ lattice was constructed through {\sc ism} iterations
which are not shown. Although $L =20$ is a rather small value
for \DMRG calculations the data in table~\ref{densitymatrices}
illustrate a typical behaviour which
is found for larger systems as well.

\begin{table}
\caption{Gaps $\Gamma=E_1(p,L)$ for the branching-fusing model eq. (\ref{bf})
with $p=0.84$ and $m=32$ states kept
as found from the iterations of the finite system method using the
density matrices $\rho^{[1]}_1$ and $\rho^{[2]}_1$.
The total length of the chain is kept fixed to $L_l + L_r = 20$, where
$L_l$ and $L_r$ indicate the lengths of the left and right parts, which
vary during the application of the finite system method.}
\vskip 0.2truecm
\begin{tabular}{|rr|c|c|}
\hline
$L_l$ &$L_r$&  $\Gamma$ from $\rho^{[2]}_1$ & $\Gamma$
from $\rho^{[1]}_1$ \\
\hline
 10 & 10  &  0.0109823881100  & 0.0109738583392 \\
 11 &  9  &  0.0109794248185  & 0.0109738587466 \\
 12 &  8  &  0.0109784192015  & 0.0109738588520 \\
 13 &  7  &  0.0109781170856  & 0.0109738588725 \\
 14 &  6  &  0.0109781024661  & 0.0109738588779 \\
 13 &  7  &  0.0109781050213  & 0.0109738588777 \\
 12 &  8  &  0.0109781103648  & 0.0109738588780 \\
 11 &  9  &  0.0109781100799  & 0.0109738588781 \\
 10 & 10  &  0.0109738162569  & 0.0109738594171 \\
 11 & 11  &  0.0109738097605  & 0.0109738594170 \\
 12 & 12  &  0.0109738065720  & 0.0109738594168 \\
 13 &  7  &  0.0109738036253  & 0.0109738594168 \\
 14 &  6  &  0.0109738015566  & 0.0109738594168 \\
 13 &  7  &  0.0109738011287  & 0.0109738594168 \\
 12 &  8  &  0.0109738036758  & 0.0109738594169 \\
 11 &  9  &  0.0109738061907  & 0.0109738594168 \\
 10 & 10  &  0.0109737989313  & 0.0109738594171 \\
\hline
\end{tabular}
\label{densitymatrices}
\end{table}

The first two columns in table~\ref{densitymatrices} show the lengths $L_l$
and $L_r$ of the right and left parts, respectively,
used to form the system of total length $L=20$.
The third and forth columns show the value of the gap $\Gamma:=E_1(p,L)$,
obtained from the density matrices $\rho^{[1]}_1$ and  $\rho^{[2]}_1$, as
defined in (\ref{densitymat}) and (\ref{densitymatright}).
In both cases the iterations during the \FSM improve the estimation
of the gap as it can be seen from the general trend of convergence.
Also, it is apparent that consideration of the symmetric partition
$L_{l}=L_{r}=L/2=10$ achieves a particularly large increase in precision.
In practice, this partition should be used for the final estimates from the
\DMRG algorithm.
Both estimates of the gap are quite close to each other, however
the density matrix (\ref{densitymat}) provides the more stable
results, with a convergence up to the 11th digit.

Although the density matrices are constructed only from the first
excited state as a target state,
we found that also the estimates of the ground state
energy are rather accurate i.e. $E_0(p,L) \approx 10^{-8}$ for
the density matrix $\rho^{[1]}_1$  and $E_0(p,L) \approx 10^{-4}$
for $\rho^{[2]}_1$. Both values are close to the exact result $E_0(p,L)
= 0$, and show again that the density matrix $\rho^{[1]}_1$ not only shows
better convergence for the targeted first excited state, but also a much more
accurate value for $E_0(p,L)$ which was not targeted. This is a practically
important remark for the following reason. Suppose one wants to find the
first excited state. In cases with a very small gap
$\Gamma$, there might be a spurious interchange with the ground state if the
the error of the \DMRG is larger than the true value of $\Gamma$.

\begin{table*}[h]
\caption{Gaps $\Gamma=E_1(p,L)$ for the branching-fusing model eq. (\ref{bf})
for $p=0.8403578$ and with $m=32$ states kept
as found from the iterations of the \FSM using the density matrices
$\rho^{[3]}$ and $\rho^{[1]}$.
$L_l$ and $L_r$ indicate the lengths of the left and
right parts, which vary during the application of the
finite system method. The last column was calculated 
using 30 digit
precision arithmetic, yielding a result free of diagonalization
inaccuracies.}
\vskip 0.2truecm
\begin{tabular}{|rr|c|c|c|}
\hline
$L_l$ &$L_r$&  $\Gamma$ from $\rho^{[3]}$ & $\Gamma$ from $\rho^{[1]}$ &
$\Gamma$ from $\rho^{[1]}$, high prec.\ \\
\hline
 6 &  6 &  0.0211852795111 & 0.0211852795111  & 0.0211852795111  \\
 7 &  7 &  0.0173940529006 & 0.0173940538620  & 0.0173940538302  \\
 8 &  8 &  0.0146003381516 & 0.0146003960454  & 0.0146003961355  \\
 9 &  7 &  0.0146004093819 & 0.0146003960889  & 0.0146003961161  \\
10 &  6 &  0.0146006193716 & 0.0146003961644  & 0.0146003961865  \\
 9 &  7 &  0.0146004129822 & 0.0146003961642  & 0.0146003961866  \\
 8 &  8 &  0.0146004043568 & 0.0146003962502  & 0.0146003962379  \\
 7 &  9 &  0.0146004373237 & 0.0146003963897  & 0.0146003962378  \\
 6 & 10 &  0.0145993594105 & 0.0146003962420  & 0.0146003962377  \\
 7 &  9 &  0.0146004306753 & 0.0146003961645  & 0.0146003962378  \\
 8 &  8 &  0.0146004003523 & 0.0146003961420  & 0.0146003962379  \\
 9 &  9 &  0.0124729781067 & 0.0124729862108  & 0.0124729862559  \\
10 &  8 &  0.0124730375718 & 0.0124729861363  & 0.0124729862333  \\
11 &  7 &  0.0124731467646 & 0.0124729861847  & 0.0124729862262  \\
12 &  6 &  0.0124732937813 & 0.0124729861234  & 0.0124729862244  \\
11 &  7 &  0.0124731477916 & 0.0124729861816  & 0.0124729862251  \\
10 &  8 &  0.0124730400858 & 0.0124729861765  & 0.0124729862258  \\
 9 &  9 &  0.0124730036975 & 0.0124729869859  & 0.0124729861961  \\
\hline
\end{tabular}
\label{mixedunmixed}
\end{table*}

Targeting simultanenously the ground and first excited states, i.e. using
density matrices of the type $(\rho^{[k]}_0 + \rho^{[k]}_1)/2$,
generally leads to an improvement of the value of $E_0(p,L)$ but
worsens the convergence
of the first excited state $E_1(p,L)$ during the \FSM iterations.
Still the behaviour of numerical convergence is analogous to that found
for targeting with a single state only.
Table \ref{mixedunmixed} shows the \DMRG calculation, first the {\sc ism}
starting from $L=12$ and then the {\sc fsm}
iterations for the branching-fusing model defined in (\ref{bf})
for chains of lengths up to $L=18$, with $m=32$ states kept and with
$p=0.8403578$, very close to the
the critical point. Here, both ground and first excited
state were targeted at the same time, using both the ``mixed'' density matrix
$\rho^{[3]}$ and the unmixed density matrix $\rho^{[1]}$
(with the shorthand notation $\rho^{[k]}:=(\rho^{[k]}_0 + \rho^{[k]}_1)/2$).
To estimate
the effect of numerical inaccuracies in the diagonalization routines, the
calculations using $\rho^{[1]}$ were redone using extended precision
arithmetic, with 30 instead of 14 mantissa digits. That is clearly enough
to preclude the possibility of
arithmetic errors to the quoted numerical precision. We observe that the choice
of the unmixed density matrix is clearly superior, unlike the case of
transfer matrix {\sc dmrg}.

Table \ref{diffmixedunmixed} shows the finite system method ({\sc fsm})
iterations for the diffusion-annihilation model (\ref{da})
for chains with up to $L=18$ sites, with $m=32$ states kept
and $p=1$. Data were obtained by targeting the third eigenstate
(lowest excited state), using both $\rho^{[3]}$ and $\rho^{[1]}$,
and are compared with the exact result (see the following section).
The choice of
the symmetric density matrix is clearly superior over against the unsymmetric
choice, by several digits of precision, even for the small number
of \DMRG iterations performed; the difference in precision increases
even further for longer systems.

\begin{table*}
\caption{Gaps $\Gamma=E_2(p,L)$ for the 
diffusion-annihilation
model eq. (\ref{da}) for $p=1$ and with $m=32$ states kept
as found from the iterations of the \FSM using the density matrices
$\rho^{[3]}$ and $\rho^{[1]}$.
$L_l$ and $L_r$ indicate the lengths of the left and
right parts, which vary during the application of the
finite system method. The last column gives the exact result.
For total length 18, only the three results for equal lengths
are given.}
\vskip 0.2truecm
\begin{tabular}{|rr|c|c|c|}
\hline
$L_l$ &$L_r$&  $\Gamma$ from $\rho^{[3]}$ & $\Gamma$ from $\rho^{[1]}$ &
exact \\
\hline
 6 &  6 &  0.0581163650912 & 0.0581163651488  & 0.0581163651479 \\
 7 &  7 &  0.0437254502992 & 0.0437047728086  & 0.0437047985324 \\
 8 &  8 &  0.0340666763278 & 0.0340537551881  & 0.0340538006322 \\
 9 &  7 &  0.0340537973881 & 0.0340537521644  & \\
10 &  6 &  0.0340538348398 & 0.0340537779114  & \\
 9 &  7 &  0.0340437188607 & 0.0340537779110  & \\
 8 &  8 &  0.0340537940475 & 0.0340538006303  & 0.0340538006322 \\
 7 &  9 &  0.0340537902152 & 0.0340538006306  & \\
 6 & 10 &  0.0340537976329 & 0.0340538006308  & \\
 7 &  9 &  0.0340537982237 & 0.0340538006312  & \\
 8 &  8 &  0.0340537985217 & 0.0340538006323  & 0.0340538006322 \\
 9 &  9 &  0.0272774311113 & 0.0272774426890  & 0.0272773931946 \\
 9 &  9 &  0.0272773902380 & 0.0272773931864  & 0.0272773931946 \\
 9 &  9 &  0.0272773945078 & 0.0272773931999  & 0.0272773931946 \\
\hline
\end{tabular}
\label{diffmixedunmixed}
\end{table*}

\section{Finite-size scaling of the relaxation time}
\label{sec:fss}

We want to test the accuracy of the \DMRG method applied to the
critical region of reaction-diffusion processes. The main interest of this
study is not so much to obtain an extremely precise value for some exponent
through a huge computational effort, but rather to get some insight into the
generic behaviour of the \DMRG in this kind of application.

\subsection{Diffusion-annihilation}

We analyse the diffusion-annihilation model
defined in (\ref{da}). The ground state is twofold degenerate
$E_0 (p,L) = E_1 (p,L) =0$, even for $L$ finite,
since the reaction $AA \to \emptyset \emptyset$
reduces the number of particles and thus there
are two stationary states: the empty lattice $|\emptyset
\emptyset \emptyset \ldots \emptyset \rangle$ and a state obtained
from the following combination of one-particle states: $| A \emptyset
\emptyset \ldots \rangle + | \emptyset A \emptyset \ldots \rangle +
\dots + | \emptyset \emptyset \emptyset \ldots A \rangle$.

The gap is known exactly (see \cite{Henk97} and references therein)
\be
\Gamma (p,L) = E_2 (p,L) = 2 p \left( 1 - \cos \frac{\pi}{L+1}
\right)
\ee
In the thermodynamic limit $L \to \infty$ one has $\Gamma (p,L) \sim
1/L^2$, i.e. a dynamical exponent $\theta = 2$.

\begin{figure}[ht]
\epsfxsize=80mm
\epsffile{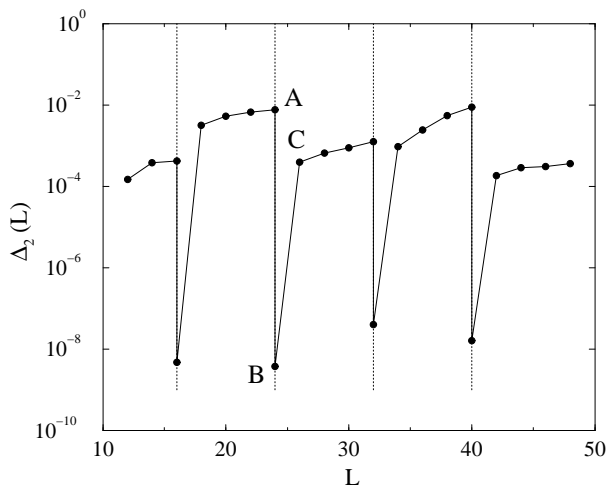}
\caption{Relative error on the gap $\Gamma (p,L)$ of diffusion-annihilation
with $p= 0.5$ as a function of the system length $L$. For $L = 16$, $24$,
$32$ and $40$ (marked by the dotted vertical lines) we report the two values
of the errors before and after the application of the finite
system method (see text).}
\label{accuracy}
\end{figure}

Figure \ref{accuracy} shows a plot of the relative error:
\be
\Delta_2 (L) := \left | \frac{E_2^{\rm DMRG} (p,L) - E_2 (p,L)}
{E_2 (p,L)} \right |
\ee
The calculation is done following the method described at the end of
section \ref{sec:dmrg}: The finite system method sweeps are applied
for chains of lengths $L = 16$, $24$, $32$ and $40$ during the same run.
For these sizes, in the figure two values
of $\Delta_2 (L)$ are given,
which are obtained at the beginning and at the end of
the application of the {\sc fsm}. We always observed a strong decrease
of the error during the \FSM iterations. For example, for $L = 24$
the point $A$ in figure~\ref{accuracy} corresponds to a relative
error equal to $10^{-2}$ and was reached before the application of the \FSM,
while after the \FSM iterations, we are at $B$, where the error drops to
$10^{-9}$. Notice that a single \DMRG step with the {\sc ism}
from $L=24$ (B) to $L=26$ (C) corresponds to a rather big increase of
the error from $10^{-9}$ to $10^{-4}$.

We conclude that for critical systems, as in the case at hand,
the use of the \FSM leads to a remarkable increase in the accuracy,
even for chains of small lengths. That is
in general not the case for \DMRG applied to symmetric problems.

Furthermore, the additional numerical precision gained from the \FSM with
respect to the {\sc ism} used alone is needed if precise information on the
critical parameters is desired. To illustrate this, consider the numerical
calculation of the dynamical exponent $\theta$. Finite-size estimates can be
obtained from $\theta(L)=-\ln(\Gamma(L+2)/\Gamma(L))/\ln((L+2)/L)$ and are
listed in table~\ref{tab:theta}.
\begin{table}
\caption{Finite-size estimates of the exponent $\theta$ for the model (\ref{da})
as a function of the system size $L$. \label{tab:theta}}
\begin{tabular}{|cc|} \hline
$L$ & $\theta(L)$ \\ \hline
10 & 1.8219 \\
18 & 1.8959 \\
26 & 1.9265 \\
34 & 1.9432 \\
42 & 1.9537 \\
50 & 1.9609 \\
58 & 1.9662 \\ \hline
\end{tabular}
\end{table}
While the sequence clearly converges toward the exact value $\theta(\infty)=2$,
it is also apparent that the finite-size data themselves are, even for the
relatively large sizes used here, still quite far from the $L\to\infty$ limit.
In conclusion, it is not possible to simply take some value $\theta(L)$
calculated on a large lattice to be a reliable estimate for the
exponent $\theta$. Rather, a careful extrapolation of the data towards their
$L\to\infty$ limit must be performed, as will be further discussed below.
However, finite-lattice extrapolations are only possible if the data are
accurate to at least 6 or 7 digits. 
Since even in the simple model (\ref{da})
finite-size effects are considerable, this should be expected to hold to an
even larger extent in more complicated systems.

The data in figure~\ref{accuracy} were obtained from the
density matrix $\rho^{[2]}_2$, defined in (\ref{densitymatright}),
with $m=32$ states kept. At least for the examples studied in this paper,
we found that the numerical accuracy cannot be further improved by
increasing $m$. This is so since the eigenvalues $\omega_i$ of the density
matrix vanish so rapidly with increasing $i$ that the truncation error
is of the same order of magnitude as the numerical errors from the other
parts of the calculation.

For systems of sizes larger than $L \approx 50$, our results become
numerically unstable. This might be due to the fact that the gap vanishes
rather rapidly in this example ($\theta = 2$). On the other hand,
it is well-known\cite{ostlund} that even the symmetric \DMRG cannot describe
critical systems in the $L \to \infty$ limit, where the gap becomes too small.

\subsection{Branching-fusing}

The branching-fusing model defined by the rates (\ref{bf}) is critical
at $p = p_c$, the value of which is not known exactly.
This critical point can be extracted from finite-size data
of the gap, using the following scaling form, valid in the vicinity
of $p_c$
\be
\Gamma (p,L) = L^{-\theta} G \left( |p - p_c| L^{1/\nu_\perp}
\right)
\label{scalgap}
\ee
where $G(x)$ is a scaling function.
Form a comparison of the gaps of three consecutive sizes, say $L-2$,
$L$ and $L+2$ one identifies $p_c(L)$ as the value of $p$ for
which the equation
\beqn
\frac{\log\left[\Gamma(p,L+2)/\Gamma(p,L)\right]}
{\log[L/(L+2)]} = \frac{\log\left[\Gamma(p,L)/
\Gamma(p,L-2)\right]}{\log[(L-2)/L]} \nonumber \\
= \theta(L)\quad
\label{sizescal}
\eeqn
is satisfied. In addition, an estimate $\theta(L)$ for the exponent $\theta$
is obtained \cite{Doma84}.

The gaps $\Gamma(p,L)$ have been calculated for chains of lengths up to $L
\approx 60$, usually with $m=32$ states kept. As before, we find that a larger
value of $m$ does not improve the results, since the density matrix eigenvalues
$\omega_i$ fall off rapidly with $i$ and typical truncation
errors for $m=32$ are of the order $\epsilon \approx 10^{-15}$.
On the other hand, the use of more than 64-bit arithmetic does
improve precision, which means that arithmetic precision is the limiting
factor here. Since for an accurate extrapolation, we need precise finite-lattice
data we merely considered chains of lengths $L \leq 30$.
These data are already sufficient to our goal to investigate the generic
behaviour of the \DMRG applied to the calculation of critical parameters.

\begin{table*}[hb]
\caption{
Finite-size estimates of critical point $p_c$ and of various
exponents for the branching-fusing model, obtained from the gap $\Gamma$,
{}from the density profile $n(l)$ defined in (\ref{n0}) 
and from the scaling of $N(l)$ as defined in (\ref{n1}).
The last row shows the $L\to\infty$ limit obtained
from \BST extrapolation. The numbers in brackets give the estimated
uncertainties in the last digit.}
\begin{tabular}{|c|ccc|cc|cc|}
\hline
 & \multicolumn{3}{c|}{from $\Gamma$} & \multicolumn{2}{c|}{from $n(l)$} &
 \multicolumn{2}{c|}{from $N(l)$} \\ \hline
$L$ & $ p_c(L) $  &  $\theta(L)$ & $\zeta(L)$ &
$ p_c(L)$ & $\beta/\nu_\perp(L)$ &
$\beta/\nu_\perp(L)$ & $\beta_1/\nu_\perp(L)$ \\
\hline
 10  & 0.815486295 & 0.830071389  & 0.177917024 &  &  &  & \\
 12  & 0.822241704 & 0.923515450  & 0.248868030 &  &
& 0.211498060 & 0.524156106 \\
 14  & 0.826556808 & 0.996672190  & 0.303005003 &  &
& 0.214641534 & 0.540022433 \\
 16  & 0.829477408 & 1.055258740  & 0.345372694 & 0.844595690 & 0.174469664
& 0.217449273 & 0.552833181 \\
 18  & 0.831547147 & 1.103159519  & 0.379339662 &  &
& 0.219928645 & 0.563377349 \\
 20  & 0.833068754 & 1.143030157  & 0.406998719 & 0.843578941 & 0.183190533
& 0.222113618 & 0.572197559 \\
 22  & 0.834221223 & 1.176727177  & 0.430122954 &  &
& 0.224042959 & 0.579677788 \\
 24  & 0.835115836 & 1.205580740  & 0.449620922 & 0.842813911 & 0.191175959
& 0.225753163 & 0.586096418 \\
 26  & 0.835824726 & 1.230565614  & 0.466327849 &  &
& 0.227274607 & 0.591673398 \\
 28  & 0.836396350 & 1.252411806  & 0.480376056 & 0.842276687 & 0.197772094
& 0.228648854 & 0.596559909 \\
 30  &             &              &             &             &
& 0.229893751 & 0.600879333 \\
 32  &             &              &             & 0.841894149 & 0.203168874
& 0.230988495 & 0.604833883 \\
 34  &             &              &             &             &
& 0.231983300 & 0.608060181 \\
 36  &             &              &             & 0.841617080 & 0.207584161
& 0.232890344 & 0.611490572 \\
 40  &             &              &             & 0.841415905 & 0.211176020
&  & \\
\hline
$\infty$ & 0.84036(1) & 1.580(1) & 0.66(2) &  0.8406(3) & 0.24(1) & 0.249(3)
& 0.667(2) \\
\hline
\end{tabular}
\label{summaryext}
\end{table*}

\begin{table*}[hb]
\caption{\BST extrapolation table for the critical
point $p_c$. The first column are the ``raw'' data obtained from
$\Gamma$ (we used $L = 16$,
$18$, $20$, \ldots $28$) and the following columns are obtained by repeated
application of the \BST transformation, with $\omega = 1.8475$.
\label{bstgap}
}
\begin{tabular}{|ccccccc|} \hline
 0.829477408 & 0.840171029 & 0.840328212 & 0.840352241 & 0.840366202
 & 0.840357906 & 0.840357626 \\
 0.831547147 & 0.840223500 & 0.840337691 & 0.840357304 & 0.840358276
 & 0.840357764 & \\
 0.833068754 & 0.840258488 & 0.840344750 & 0.840357765 & 0.840357763
 & & \\
 0.834221223 & 0.840282908 & 0.840349176 & 0.840357764
 & & & \\
 0.835115837 & 0.840300341 & 0.840351951
 & & & & \\
 0.835824726 & 0.840313022
 & & & & & \\
 0.836396350
 & & & & & &
\\ \hline
\end{tabular}
\end{table*}

In table~\ref{summaryext}, we display the finite-size data for the critical
point $p_c$ and several critical exponents which we are going to analyse.

We begin by estimating $p_c$. By looking at the data, we see that the
differences $p_c(L+2)-p_c(L)$ decrease only slowly with increasing $L$. This
already implies that the limit $p_c=p_c(\infty)$ must be quite far away from
the finite-lattice data. In fact, this situation had already been encountered
before for periodic boundary conditions, when studying the finite-size scaling
of Reggeon field theory, which is also in the universality class of
directed percolation. There, it was shown \cite{Henk90} that the $L\to\infty$
extrapolation may be reliably carried out with the \BST algorithm
\cite{BST,Henk88}.
This algorithm transforms a `logarithmically converging' (e.g. \cite{Henk99})
sequence into another one which is expected to converge faster. It involves
a free parameter $\omega$ which roughly measures the effective leading
correction exponent and is chosen to achieve optimum convergence. Comparison
of the behaviour of the \BST algorithm and several alternative extrapolation
schemes applied to finite-size data \cite{Henk88,Barb82} has shown that in
the generic case when the correction exponent has a non-integer value, the \BST
algorithm is the most reliable of the schemes presently available.

The extrapolation procedure is illustrated in table~\ref{bstgap}.
The first column gives the values of $p_c(L)$ which solve eq.~(\ref{sizescal}),
taken from table~\ref{summaryext}.
The subsequent columns present the convergence-accelerated sequences as found
from the \BST transformation. For the chosen value of $\omega$, stability is
found and a conservative estimate of the location of the critical point is
$p_c = 0.84036(1)$. Notice that the final value of $p_c$ is indeed quite far
from the finite-lattice data, however, the fact that the given sequence can be
made to converge well indicates that the original set of \DMRG data is fairly
accurate. The possibility of estimating $p_c$ precisely mainly depends on the
length of the sequence available for extrapolation and not so much on the
distance of the finite-size data from their $L\to\infty$ limit.

In the same way, the data obtained from (\ref{sizescal}) for the dynamical
exponent $\theta$ can be analysed. Its determination is independent of the
final estimate of $p_c$. We do not present the details, but
simply quote our result $\theta = 1.580(1)$. First of all, it is satisfactory
to see the good agreement with the value quoted in table \ref{MCexp}. Second,
we point out that even for $L=28$, the finite-lattice estimate is still some
$20\%$ away from its limit value. For comparison, we recall that for periodic
boundary conditions (in the Reggeon field theory and with an accuracy for
$\theta(\infty)$ comparable to the situation at hand) for $L=14$,
the corresponding difference is of the order of $3\%$ \cite{Henk90}.
This is a consequence of the free boundary conditions usually employed with
the {\sc dmrg}. It is remarkable that in spite of the extra difficulty presented
by the free boundary conditions, one is still capable to determine $\theta$
so precisely. In order to improve the precision, one would have to perform
the \DMRG with enhanced numerical accuracy in order to 
generate longer sequences.

Differentiating numerically eq. (\ref{scalgap}),
one finds
\be
\frac{\partial \Gamma (p,L)}{\partial p} =
L^{-\theta + 1/\nu_\perp} \,\, G^\prime \left (
|p - p_c| L^{1/\nu_\perp} \right )
\ee
which allows to estimate the exponent $\zeta = \theta - 1/\nu_\perp$.
Here, the numerical derivatives were calculated at the values of $p_c(L)$ 
given by the solutions of eq. (\ref{sizescal}).
{}From numerical extrapolation we find $\zeta = 0.66(2)$, and
with the value of $\theta$ found above we find for the
(spatial) correlation length exponent $\nu_\perp = 1.08(2)$.
This is also in good agreement with the value of table \ref{MCexp}.
It is somewhat less accurate, however, than the estimated value for $\theta$,
since its determination involves the calculation of numerical
derivatives.

\begin{figure}[ht]
\epsfxsize=80mm
\epsffile{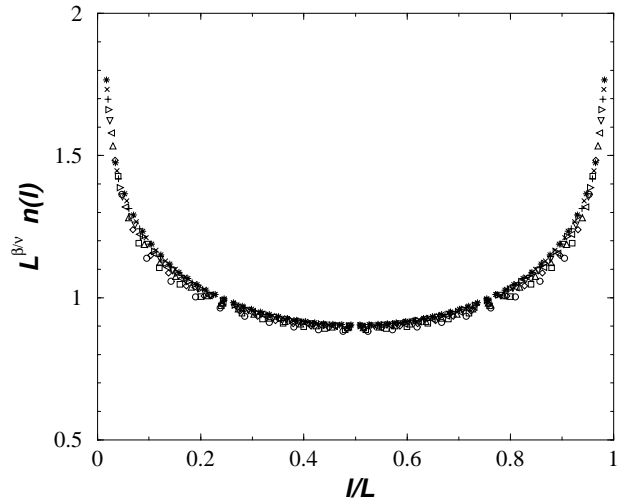}
\caption{Scaling plot of the particle density $L^{\beta/\nu_\perp} n(l)$
as a function of the scaled distance $l/L$ at the estimated critical point
$p_c = 0.84036$ and for an injection rate of $p' = 0.3$.}
\label{den0.3}
\end{figure}

\section{Density profiles in the branching - fusing
model}
\label{sec:densityprof}

\subsection{Finite-size scaling of the density profiles}

Besides the calculation of gaps, the \DMRG allows
investigating density profiles in the steady state. If $\hat{n}(l)$ is the
density operator at position $l$, the density profile is naturally calculated
from the ground states $|s\rangle = |\psi_0^{(r)}\rangle$ and
$\langle s|=\langle \psi_0^{(l)}|$, normalised as $\langle s|s\rangle =1$, as
\BEQ \label{n0}
n(l) = \langle \psi_0^{(l)} |\hat{n}(l) |\psi_0^{(r)}\rangle
\EEQ
However, for the models (\ref{da},\ref{bf}), this will be
non-vanishing only if particles are injected
at the boundaries. Therefore, we add the boundary reaction (\ref{inject})
to the branching-fusing model and
analyse the behaviour of the system as a function of the injection rate $p'$.

Adapting the finite-size scaling arguments formulated originally by Fisher and
de Gennes \cite{FdG} for equilibrium profiles, the density of particles at
the critical point $p_c$ should follow a scaling form
\be
n(l) = l^{-\beta/\nu_\perp} F(l/L)
\label{scalform}
\ee
where $0 \leq l \leq L$ denotes the position along the system, $F(x)$ is
a scaling function and $\beta$ the order parameter critical exponent.
According to (\ref{scalform}) the quantity $L^{\beta/\nu_\perp} n(l)$
depends on $l$ only through the scaling variable $l/L$.

Fig. \ref{den0.3} shows a scaling plot of $L^{\beta/\nu_\perp} n(l)$
at $p=p_c$ as determined above and for
particle injections rate $p' = 0.3$, using the value $\beta/\nu_\perp =
0.25208$, as quoted in \cite{Muno98}.
We see that the densities for systems of lengths $L = 20, 24, 28 \ldots 56$
tend to collapse nicely onto a
single curve. Nevertheless, notice that finite-size corrections are
quite large, in particular, they are larger than encountered for example in
Ising model calculations.

We now try to find $p_c$ and $\beta/\nu_{\perp}$ from our finite-lattice data.
Consider the point $l=L/2$ in the middle of the chain. The central density
$n(L/2;p)$ should obey in the vicinity of the critical point the scaling law
\be
n(L/2;p) = L^{-\beta / \nu_\perp} H \left( |p - p_c| L^{1/\nu_\perp} \right)
\ee
where $H(x)$ is a scaling function. In analogy to the analysis of the gap,
we can compute finite-size estimates for the critical point $p_c$ and the
exponent $\beta/\nu_{\perp}$ by replacing in eq.~(\ref{sizescal}) $\Gamma$ by
$n$ and $\theta$ by $\beta/\nu_{\perp}$, with the results
listed in table~\ref{summaryext}, for $p^\prime = 0.3$.

Table \ref{bstprof} shows the \BST
extrapolation table for the values of $p_c (L)$, for which we used strips
of widths up to $L = 40$. The extrapolation yields $p_c = 0.8406(3)$.
While this agrees quite well with the earlier estimate $p_c=0.84036$ found from
the gap $\Gamma$, the apparent difference between the two results gives an
{\it a posteriori} assessment of the reliability of the extrapolation procedure.

In addition, we observe that the sequence of values for $p_c$ obtained from
$\Gamma$ increases with increasing $L$, while the sequence obtained from $n$
decreases. Although the raw data in table~\ref{bstgap} are much farther away
from the $L\rar\infty$ limit of $p_c$ than those in table~\ref{bstprof},
the extrapolation in the former case can be carried out to a higher degree
of precision than in the latter case. This illustrates once more the importance
of a careful finite-lattice extrapolation procedure when trying to extract
precise parameter values from lattice calculations.

In a similar fashion as done for the exponent $\theta$, we have estimated
the ratio $\beta/\nu_\perp = 0.24(1)$.
These results, both for $p_c$ and for the exponents, are less accurate than
those obtained from the finite-size scaling analysis of the gap, indicating
that the numerical accuracy of eigenvectors of the non-symmetric Hamiltonian
$H$ for the branching-fusing model is inferior to that of its eigenvalues.

\begin{table*}[htb]
\caption{\BST extrapolation table for the critical $p_c$
from the scaling of the density of particles $n(L/2)$ in the middle
of the system, where sizes from $L = 16$ up to $L = 40$ were used, with
$\omega = 2.316$.\label{bstprof}}
\begin{tabular}{|ccccccc|} \hline
 0.844595690 & 0.842080802 & 0.840373866 & 0.840653412 & 0.840568730 &
 0.840563673 & 0.840569610 \\
 0.843578941 & 0.841361648 & 0.840593970 & 0.840558615 & 0.840564091 &
 0.840556196 & \\
 0.842813911 & 0.841027230 & 0.840571147 & 0.840563201 & 0.840572164 &
 & \\
 0.842276687 & 0.840840423 & 0.840566363 & 0.840610429 &
 & & \\
 0.841894149 & 0.840734840 & 0.840594342 &
 & & & \\
 0.841617080 & 0.840683057 &
 & & & & \\
 0.841414666 &
 & & & & &
 \\ \hline
\end{tabular}
\end{table*}

\subsection{The limit $p^\prime \to 0$}

We now discuss the consequences of varying the boundary injection rate $p'$,
in particular the limit $p'\to 0$.

\begin{figure}[b]
\epsfxsize=80mm
\epsffile{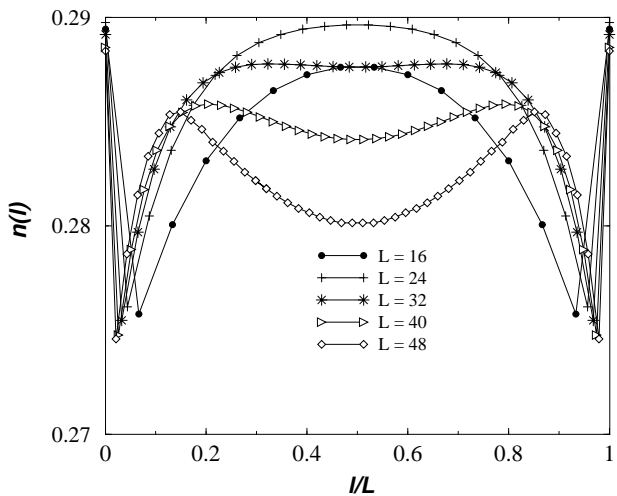}
\caption{Particle density $n(l)$ at the estimated critical point $p_c =
0.84036$ for an injection rate $p' = 0.03$ plotted as function of
$l/L$.}
\label{den0.03}
\end{figure}

Fig. \ref{den0.03} shows the particle density $n(l)$ as function
of the scaled variable $l/L$ for an injection rate $p' = 0.03$,
$L = 16$, $24$, $32$, $40$ and $48$ at the estimated critical point 
$p = p_c$.
The behaviour is now completely different with respect to the profiles
with $p'=0.3$. Apart from the two boundary sites (corresponding to $l=
0$ and $l = L$), where the density of particles is weakly dependent on
the system size $L$, we see that starting from the edges the density
increases towards the bulk. For small systems the maximum of the density 
is found in the middle of the chain, while for systems large enough two 
maxima are located at a distance $l_{\rm max} \approx 7$ lattice spacings 
from the edges (we find that $l_{\rm max}$ is independent of $L$ for 
sufficiently long chains).

This phenomenon is rather counterintuitive since one would expect that
the density of particles as a function of $l$ should decay monotonically 
starting from the edges and moving towards the center, as seen in 
figure~\ref{den0.3} for $p^\prime = 0.3$.
The effect observed here actually has a counterpart in the magnetization
profiles of equilibrium spin systems in the presence of a weak surface
magnetic field \cite{weakh1}. This effect in turn was found and explained by
appealing to the universal short-time critical dynamics involving the
so-called slip exponent \cite{janssen}. Non-monotonous profiles of this
kind are a true fluctuation effect and cannot be explained on the level
of a mean-field approximation, see \cite{weakh1,janssen,diehl,fredrik} and
appendix~\ref{sec:appendixc}.
It is known, for instance for the Ising model at its critical point, that
a weak surface field $h_1$ induces a {\it macroscopic} scale length $l_1$
such that up to a distance $l \approx l_1$ from the surface, the magnetization
is increasing with $l$. It starts to decrease again as $l > l_1$, with the
asymptotic behaviour $\sim l^{-\beta/\nu}$ for large distances (in the Ising
model, $\nu_\perp = \nu_\| = \nu$) \cite{andrzej}.

To describe non-monotonic densities of particles, it is necessary to modify
the scaling form (\ref{scalform}) to include a term depending on the
boundary reaction rate $p^\prime$, following the standard
theory of boundary critical phenomena at the ordinary transition,
see \cite{weakh1,diehl}. The rate $p^\prime$ will enter into the particle
density in the
form of a scaling variable $\lambda = {p^\prime} l^{x_1}$, where $x_1$ is
some scaling dimension. One expects the scaling form
\be
n(l,p^\prime) = l^{-\beta/\nu_{\perp}} {\cal F} (l/L, {p^\prime} l^{x_1})
\label{scalpprime}
\ee
Following the same analysis as
presented in \cite{weakh1,diehl}, one finds that $x_1 = \beta_1 / \nu_\perp$,
with $\beta_1$ the (ordinary) order parameter surface exponent.
This exponent has been calculated for directed percolation, see 
table~\ref{MCexp}, yielding $x_1 \simeq 0.669$.

For large $\lambda$, one should recover from (\ref{scalpprime})
the scaling form given in eq. (\ref{scalform}), thus
\be
\lim_{\lambda \to \infty} {\cal F} (l/L, \lambda ) = F(l/L)
\ee
More interesting, for the present case, is the other limit where
$\lambda\to 0$. In absence of particle injection ($p^\prime = 0$),
the particle density $n$ vanishes, which implies ${\cal F} (x,
\lambda = 0) = 0$. At small $p^\prime$, it is natural to assume that
the density varies linearly with $p^\prime$, leading to the condition
\be
{\cal F} (l/L, \lambda ) \sim \lambda
\label{lambdaasmall}
\ee
valid in the range, say, $0 \leq \lambda < \lambda_0$. Notice that
$\lambda_0$ defines a length scale $l_1 = (\lambda_0/p^\prime)^{1/x_1}$.
Therefore, in the $L \ll l_1$ limit, one finds
\be
n(l,L) = l^{\rho^\prime} \widetilde{{\cal F}}(l/L)
\label{pprimetozero}
\ee
obtained by inserting the limiting value (\ref{lambdaasmall}) in eq.
(\ref{scalpprime}) and where $\rho^\prime = x_1 - \beta/\nu_\perp =
(\beta_1 - \beta)/\nu_\perp$, in complete analogy with \cite{weakh1}.

{}From the numerical values quoted in table~\ref{MCexp}, we find $\rho^\prime
\simeq 0.416$, the positivity of which explains the increase of the
profiles with $l$ close to the boundary. Actually, the particle density
with $p'=0.03$ shown in Fig. \ref{den0.03} corresponds
neither to the scaling regime (\ref{scalform}) nor to (\ref{pprimetozero}),
but to an intermediate situation.

\begin{figure}[b]
\epsfxsize=80mm
\epsffile{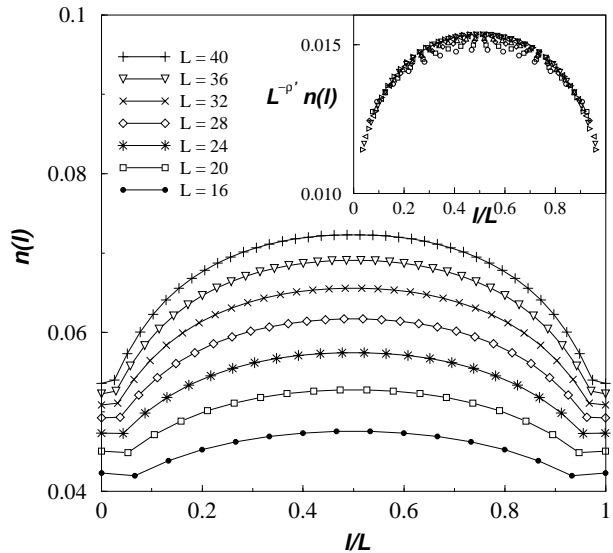}
\caption{Density of particles at $p_c$ and for $p^\prime = 0.002$.
The inset shows the collapse of the scaled densities onto a single curve as
expected from  eq.~(\ref{pprimetozero}).}
\label{figd0.002}
\end{figure}

In figure~\ref{figd0.002}, we display the critical particle density for
$L = 16$, $20$, \ldots $40$ but with a decreased injection rate
$p^\prime = 0.002$. Profiles are monotonously increasing from the edges and
there is no sign of an inflection of the curves as seen in Fig. \ref{den0.03}.
The inflection point should be found at a distance from the edges 
$l_{\rm max} (p^\prime = 0.002)$ given by the relation:
\begin{displaymath}
l_{\rm max} (p^\prime = 0.002) = l_{\rm max} (p^\prime = 0.03)
\left( \frac{0.03}{0.002} \right) ^{1/x_1} \approx 400
\label{lmaxscal}
\end{displaymath}
in units of the lattice constant, 
and where we have used $l_{\rm max} (p^\prime = 0.03) \approx 7$
as estimated from the profiles of Fig. \ref{den0.03}.
Thus, we expect that for $p^\prime = 0.002$, at about $400$
lattice spacings from the surface the density profiles should show
a maximum and then start decreasing again. This effect should 
be noticeable only for systems of sizes $L > 2 l_{\rm max} \approx 800$, 
well beyond the range of sizes shown in figure \ref{figd0.002}.

The inset of figure~\ref{figd0.002} shows the scaled profiles
$L^{-\rho^\prime} n(l)$ plotted as a function of the scaled distance
$l/L$ (we have used the expected value of $\rho^\prime = 0.416$ for 
directed percolation and removed the edge sites $l = 0$ and $l =
L$ from the plot). If the scaling assumptions leading to 
eq.~(\ref{pprimetozero}) were correct, all the profiles for different 
values of $L$ should collapse into a single curve. 
Indeed, we find a trend towards a data collapse, although finite-size
corrections are rather strong, especially in the middle of the chain, while
they are weaker at the edges. It is remarkable, that the scaling law
(\ref{pprimetozero}), whose derivation involves the scaling close to the
boundary, should be valid for the entire finite system, at least for the
lattice sizes under consideration here.
On the basis of these observations, one may conclude that the \DMRG data
for the profiles agree with the results of the scaling theory also in
the weak injection rate regime.

\subsection{Profiles from matrix elements}

So far, the calculations of the density profiles used the straightforward form
(\ref{n0}) and then tried to perform the limit $p'\to 0$ numerically, which in
principle should be taken only {\em after} the $L\to\infty$ limit has been
carried out. Alternatively, we may rely on an analogy with the calculation of
the order parameter of equilibrium spin systems which avoids this cumbersome
double limit \cite{Hame82}, see also \cite{Henk99}.
If $|\psi_{1}^{(r)}\rangle$ and $\langle\psi_{1}^{(l)}|$ are the first excited
eigenstates of $H$, consider \cite{NormN}
\be
N (l) := \langle \psi_1^{(l)} | {\hat n} (l) | \psi_1^{(r)} \rangle
\label{n1}
\ee
where ${\hat n} (l)$ is the density operator at position $l$ and where we have
simply set $p'=0$. Although $N(l)$ is not directly related to the more
``physical'' density $n(l)$, it offers the distinct advantage that it is
non-vanishing even for $p'=0$ and furthermore displays the same scaling
behaviour as expected for $n(l)$. Since in the percolating phase, the
first gap $\Gamma$ is exponentially small for $L$ large, one might expect
to find for $N(l)$ the same scaling behaviour as for $n(l)$. Indeed, 
as we shall see, the following finite-size
scaling behaviour holds for $N(l)$ at $p=p_c$: 
$N(l) \sim L^{-\beta/\nu_{\perp}}$
for sites deep in the bulk, i.e. $l/L \simeq 1/2$ and
$N(l) \sim L^{-\beta_1/\nu_{\perp}}$ for boundary sites, i.e. $l=0$ or
$l=L$; with the same exponent values as found before. 

\begin{figure}[b]
\epsfxsize=80mm
\epsffile{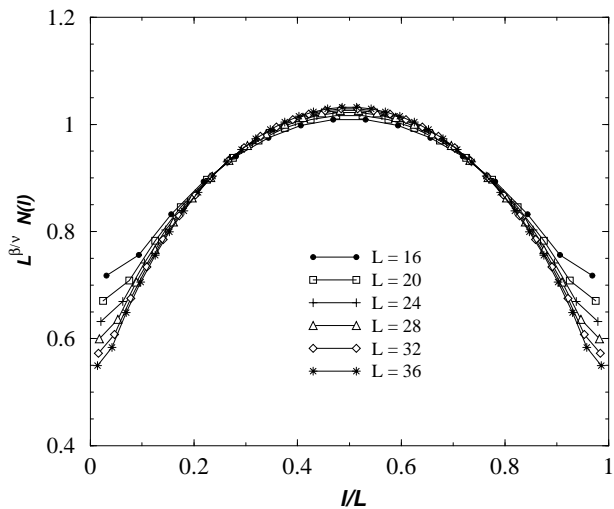}
\caption{Plot of the scaled quantities $L^{\beta / \nu_\perp} N (l)$
as function of the scaled distance $l/L$ for $L = 16, \ldots$ at the
critical point $p=p_c$.}
\label{mat11}
\end{figure}

Figure \ref{mat11} shows a scaling plot of $L^{\beta/\nu_\perp} N(l)$
as a function of $l/L$, where we used the value of $\beta/\nu_\perp=0.25208$
{}from table \ref{MCexp}. The data collapse in a
satisfactory way in the bulk, but rather poorly at the surface. This is
consistent with our expectation of two different scaling regimes. It is 
also quite remarkable that the numerical data for $l = L/4$ and $l = 3 L/4$ 
show an almost perfect collapse.

We analyzed the numerical data at $p = p_c$ forming the finite-size estimates
from two successive system sizes as done in eq. (\ref{sizescal}), using the
values of $N(l)$ at $l =L/2$ and at the surface $l = 1$. The data are
listed in table~\ref{summaryext}. Extrapolation for $L\to\infty$
with the \BST method yields $\beta/\nu_\perp = 0.249(3)$ and $\beta_1/\nu_\perp
= 0.667(2)$, both in good agreement with the results quoted in
table~\ref{MCexp}. Apparently the exponents extrapolated from $N(l)$ are more
accurate than those obtained from the density profiles and allow a direct
determination of surface exponents, allowing a comparison with field-theory
methods \cite{per}. 
Again, the data for $N(l)$ might also be
used to find yet another estimate of $p_c$, but we have not carried out this
calculation.

\section{Conclusions}
\label{sec:conc}

In this paper we have investigated the properties of two reaction-
diffusion models at their critical point, by means of \DMRG techniques.
Our study demonstrates that the \DMRG is capable to treat rather
well out-of-equilibrium systems and provides a method for calculations
of critical exponents, alternative to other general methods such as
Monte Carlo simulations or series expansion techniques. The \DMRG offers the
advantages that (i) there is no critical slowing-down, (ii) the method does
not require random numbers and (iii) there is no need to make any assumptions
about the basic state around which one may expand.

Our findings may be summarised as follows.

\noindent (1) In the examples considered, the symmetric density matrix
$\rho^{[1]}$ produced the most accurate results. It is essential to have
a reliable method to diagonalize a sparse non-symmetric matrix.
The non-symmetric Lanczos and the Arnoldi algorithms used here were found to
produce results of comparable accuracy.

\noindent (2) It was enough to keep $m=32$ states. Improvements in numerical
accuracy of the \DMRG cannot be achieved by increasing $m$, but rather by
enhancing the number of digits kept beyond the conventional 64-bit arithmetics.

\noindent (3) In order to analyse the critical region, it is essential to use
the finite system method of the {\sc dmrg}. The infinite system method alone
is not sufficient.

\noindent (4) Finite-size data for critical non-equilibrium systems are
strongly affected by finite-size corrections. Because usually the \DMRG works
best for free boundary conditions, finite-size corrections should be expected
to be substantially larger than in calculations done with standard
diagonalization techniques, which usually work with periodic boundary
conditions. However, the trade-off is that from a single \DMRG calculation one 
may get both bulk and surface exponents.

\noindent (5) In studies of the scaling of profiles, the consideration of
matrix elements of the type of (\ref{n1}) may provide a better computational
tool than the physically more straightforward profiles (\ref{n0}).

\noindent (6) Precise estimates of the location of the critical point and of
critical parameters can only be obtained through a finite-sequence extrapolation
technique. That requires a long sequence of precise lattice data for several
distinct sizes, rather than data from a single large lattice.

\noindent (7) The extrapolation can be carried out reliably by the \BST
algorithm. The results for the critical exponents are in agreement with
the results for series expansion and Monte Carlo simulation,
see table~\ref{MCexp}. At present, the
\DMRG do not yet achieve the same precision as these. The precision of the
method can be improved by generating longer chains. To overcome the numerical
instabilities encountered, this will require to go beyond the usual 64-bit
arithmetics.

All in all, we conclude that the \DMRG has turned out to be a reliable
general-purpose method, also for studying non-equilibrium critical phenomena,
although it works best out of criticality.

\subsection*{Acknowledgements}

We thank P. Fr\"ojdh for fruitful correspondence and
the Centre Charles Hermite of Nancy for providing substantial 
computational time. M.H. thanks A. Honecker for a very useful discussion. 

\appendix

\section{Relation with site and bond percolation}
\label{sec:appendixa}

We recall the relationship of the reaction-diffusion processes defined in
(\ref{diff}-\ref{decoag}) with directed percolation. The latter is usually
defined in terms of local probabilities giving the state of a site at time
$t+1$ in terms of its two parent sites at time $t$
\BEQ \label{eq:A1}
\vekz{\circ\;\circ}{\bullet} = 0 \;\; ; \;\;
\vekz{\bullet\;\circ}{\bullet} = p_s p_b \;\; ; \;\;
\vekz{\bullet\;\bullet}{\bullet} = p_s p_b(2-p_b)
\EEQ
where $p_s$ and $p_b$ describe site and bond percolation, respectively. Here
$\bullet$ marks an occupied and $\circ$ an empty site. Furthermore,
$\vekz{\bullet\;\circ}{\circ} = 1 - \vekz{\bullet\;\circ}{\bullet}$ and so on.
{From} these rules, a time evolution operator $S$, acting according to
$|P(t+1)\rangle = S |P(t)\rangle$, is constructed, which is related to the
quantum Hamiltonian $H=1-S$.

In order to link the percolation parameters $p_s,p_b$ with the reaction and
diffusion rates $\alpha,\beta,\gamma,\delta$ and $D$, we observe that the
evolution of a site is according to (\ref{eq:A1}) only dependent on its parentsand independent of the state of its neighbours at the {\em same} time $t$. Thus
\BEQ
\vekz{\bullet\;\circ}{\bullet} =
\vekz{\bullet\;\circ}{\bullet\;\circ} + \vekz{\bullet\;\circ}{\bullet\;\bullet}
=\vekz{\bullet\;\circ}{\circ\;\bullet}+\vekz{\bullet\;\circ}{\bullet\;\bullet}
\EEQ
by summing over the possible states of the right or left nearest neighbours.
Similar relations hold for the other rates. From this, the following
conditions for a mapping of the process (\ref{diff}-\ref{decoag}) onto directed
percolation hold
\BEQ
\beta=1-\delta=p_s p_b \;\; ; \;\;
D = 0 \;\; ; \;\;
2\alpha + \gamma = 1 - p_s p_b (2-p_b)
\EEQ
It is easy to check that the model (\ref{bf}) cannot be exactly mapped onto
directed percolation this way,
although it still is in the same universality class.
For example, a mapping onto site percolation ($p_b=1$) requires that
$\delta=2\alpha+\gamma$, while bond percolation ($p_s=1$) is achieved for
$\delta^2=2\alpha+\gamma$.

\section{On the density matrix}
\label{sec:appendixb}

To justify the use of the density matrix
$\tr |\psi_0\rangle\langle\psi_0|$, one writes the targeted state in a
product state of states $|\alpha_l,i_l\rangle \equiv |i\rangle$ of
the ``system'' one wants to describe and $|j_r,\beta_r\rangle
\equiv |j\rangle$ of the ``environment''. From the construction of
the {\sc dmrg}, each of the bases has $mn$ elements, where $n$ is the
number of states per site. Then
\be
|\psi_0\rangle = \sum_{ij} \psi_0(i,j) |i\rangle |j\rangle
\ee
One now approximates $|\psi_0\rangle$ by $|\tilde{\psi}_0\rangle$,
with
\be
|\tilde{\psi}_0\rangle = \sum_{\alpha,j} \tilde{\psi}_0(\alpha,j)
|\alpha\rangle |j\rangle
\ee
where the sum over $\alpha$ runs over $m$ orthogonal states in the
system basis, $|\alpha\rangle = \sum_{i} u_{\alpha i} |i\rangle$.
One now demands that the approximation is optimal by minimising
\be
\| |\psi_0\rangle - |\tilde{\psi}_0\rangle \|^2 = 1 -2 \sum_{\alpha ij}
\tilde{\psi}_0^2(\alpha,j).
\ee
Making this stationary with respect to $\tilde{\psi}_0 (\alpha,j)$ gives
\linebreak
$\sum_{i} \psi_0(i,j) u_{\alpha i} = \tilde{\psi}_0 (\alpha,j)$ and thus
the global minimum of
\be
\| |\psi_0\rangle - |\tilde{\psi}_0\rangle \|^2 = 1 - \sum_{\alpha ii'}
 u_{\alpha i} \rho(i,i') u_{\alpha i'}
\label{minim}
\ee
has to be found. Here one introduces the reduced density matrix
\be
\rho(i,i') = \sum_j \psi_0(i,j)\psi_0(i',j) .
\ee
The stationary points of (\ref{minim}) are obviously given by setting
$|\alpha\rangle$ equal to the eigenvectors of the reduced density matrix
(Ritz condition). The stationary values of (\ref{minim}) are then given by
$1-\sum_{i=1}^m \lambda_i$, where $\lambda_i$ are the eigenvalues of
the density matrix ($0\leq\lambda_i\leq 1$, $\sum\lambda_i=1$).
The global minimum is obtained by choosing the $m$ eigenvectors with
the largest associated eigenvalues.

In the non-symmetric case, repeating the same argument, one
finds that the minimization of:
\be
\| |\psi_i^{(l)} \rangle - |{\tilde{\psi}}_i^{(l)} \rangle
\|^2 +
\| |\psi_i^{(r)} \rangle - |{\tilde{\psi}}_i^{(r)} \rangle
\|^2
\ee
yields as optimal basis set the eigenvectors of the density matrix
(\ref{densitymat}).

\section{Density profiles from kinetic equations}
\label{sec:appendixc}

We show that for branching-fusing (\ref{bf}),
the mean field steady-state density profile for the semi-infinite
system $x\geq 0$ with a prescribed boundary density is a monotonous function
of the distance $x$ from the boundary. If $a(x,t)$ is the mean particle
density, the kinetic equation is
\BEQ
\dot{a} = D a'' + 4 (p-1/2)\, a - 2 a^2
\EEQ
In the steady state, $\dot{a}=0$.
We write $a(x)=a_{\infty} \vph(x/\xi_{\perp})$,
where $a_{\infty}=2p-1\sim \xi_{\perp}^{-2}$ is the bulk density and
$\xi_{\perp}=\sqrt{D/(p-1/2)}$ the spatial correlation length. The profile is
\BEQ
\vph(y) =\frac{3}{2}\left(
\frac{\sqrt{(2\vph_0+1)/3\,}+\tanh(y)}{1+\sqrt{(2\vph_0+1)/3\,}\,\tanh(y)}
\right)^2 -\frac{1}{2}
\EEQ
where $\vph_0=a(0)/a_{\infty}$ is related to the boundary density. Evidently,
$\vph(y)\to 1$ as $y\to\infty$ monotonously. The above result was derived for
a semi-infinite system with $p>p_{c,{\rm MF}}=1/2$. For a finite system at
$p=1/2$, the $p$-dependence of the correlation length $\xi_{\perp}$ is traded
for a size-dependence.


\begin{thebibliography}{999}
\bibitem{ARPACK} The package can be found in ftp://caam.rice.edu
in the directory pub/people/software/ARPACK.
\bibitem{NormN} Here, the eigenvectors were normalized according to
$\| |\psi_1^{(r,l)}\rangle \|=1$.
\end{thebibliography}
\end{document}